\documentclass[%
 reprint,
 amsmath,amssymb,
 aps,
prb,
]{revtex4-2}
\usepackage{graphicx}
\usepackage{dcolumn}
\usepackage{bm}
\usepackage{siunitx}
\usepackage{kantlipsum}
\usepackage{appendix}
\usepackage{filecontents}

\begin{document}

\preprint{APS/123-QED}

\title{Electrical detection of parallel parametric amplification and attenuation in $\mathrm{Y}_3\mathrm{Fe}_5\mathrm{O}_{12}$/$\mathrm{Pt}$ bilayer disk}

\author{Geil Emdi$^{1}$}
\author{Tomosato Hioki $^{1,2}$ }
\email{tomosato.hioki@ap.t.u-tokyo.ac.jp}
\author{Koujiro Hoshi$^{1,3}$}
\author{Eiji Saitoh$^{1,2,3,4}$}
 
\affiliation{$^{1}$Department of Applied Physics,The University of Tokyo,Tokyo 113-8656,Japan}
\affiliation{$^{2}$Advanced Institute for Materials Research, Tohoku University, Sendai 980-857,Japan}
\affiliation{$^{3}$Institute for AI and Beyond, The University of Tokyo,Tokyo 113-8656, Japan}
\affiliation{$^{4}$Advanced Science Research Center, Japan Atomic Energy Agency, Tokai, Ibaraki 319-1195, Japan}





\date{\today}

\begin{abstract}
We report a systematic quantitative evaluation of parametric amplification gain of magnetization dynamics in ytirrium iron garnet ($\mathrm{Y}_3\mathrm{Fe}_5\mathrm{O}_{12}$) thin disk via a.c. spin pumping and inverse spin Hall effect. We demonstrate its signature phase-dependence where amplification and attenuation occur every $\frac{\pi}{2}$ phase shift of the input signal. The results also show the pump-power dependence of the gain that is explained well by our theoretical model. Finally, the optimal conditions for the amplification is investigated by measuring the magnetic field dependence, where we find the highest gain of 11.4 dB.
\end{abstract}

\maketitle


Degenerate parametric amplification refers to an amplification of input waves by using an excitation whose frequency is twice the resonance frequency. This amplification is phase sensitive due to the non-linearity of a medium. Owing to this phase-sensitive nature, the parametric process can selectively amplify a signal based on its phase and frequency. This selective amplification is widely used in applications such as wavelength conversion of light \cite{Loudon2000,Parker2022,Caves1987} and microwave amplification at cryogenic temperatures \cite{Aumentado2020,Aumentado2022,Yamamoto2008,Yamamoto2016}.

Spin waves, a precessional motion of magnetization in a magnetic material that propagates as waves, can undergo parametric amplification due to the strong nonlinearity from dipolar and exchange interactions \cite{Rez,Brach2016,Brach2017,Makiuchi2021,Serga2003,Melkov1999,Melkov2001}. In magnonics, which aims to realize unconventional computing systems using spin waves, the phase dependence of parametric amplification is extremely important for encoding information into the phase of magnons, a quanta of spin waves. Many paths have been proposed to utilize parametric amplification process in spin wave circuits \cite{Brach2016,Brach2017,Makiuchi2021}.
By exploiting the fact that in a parametric oscillation process, the phase of the output signal wave can be discretized into two, which can be used for probabilistic bit operation to realize optimization solvers \cite{Makiuchi2021,Hioki2021,Elyasi2022,Hioki2022}.

In this study, by measuring the parametric amplification process of spin waves using a spin current measurement method, we systematically evaluated its amplification rate, thereby clarifying the systematic behavior of the spin wave amplification gain. In the experiments, by investigating gain dependence on  the phase of the spin wave signal and the pump amplitude, we demonstrate that the spin wave amplification gain can exceed of more than an order of magnitude under optimal conditions, and we show that this behavior can be theoretically modelled by considering loss of spin waves in the magnet and electromagnetic waves in a pick up circuit.

We use an ytirrium iron garnet ($\mathrm{Y}_3\mathrm{Fe}_5\mathrm{O}_{12}$; YIG) micro-sized thin disk that is covered with platinum (Pt) thin layer as shown in Fig. 1(a). The diameter of the YIG disk is 500 $\si{\micro m}$, and the thickness is 1.4 $\si{\micro m}$. We grow the YIG layer by Liquid Phase Epitaxy (LPE) process  on top of  gadolinium gallium garnet (GGG) substrate. Then, the 10-nm-thick Pt film is sputtered on top of the YIG layer. The Pt film is used for measuring the magnetization precession in the YIG disk as a.c. voltage via the a.c. spin pumping and ISHE \cite{Wei2014,Hahn2013,Jiao2013,Weiler2014,Saitoh2006,Azevado2005,Kimura2007,Valenzuela2006,Costache2006,Mizukami2002,Kajiwara2010,Tserkovnyak2002}. The disk shape of YIG/Pt bilayer is patterned by photolithography and Ar-ion milling process. Moreover, two gold electrodes are sputtered at the edge of the Pt film. The fabricated sample is placed on top of a pump coplanar waveguide (CPW) that is short-ended, with the width of 100 $\si{\micro m}$. One of the gold electrodes is grounded, while the other is connected to an input CPW [see Fig. 1(a)]. We measure the reflection ($|S_{11}|$) spectrum of the sample from the two CPW ports and we choose the frequency of the weak input field to be 1\textit{f} = 2.15 GHz to satisfy the ferromagnetic resonance (FMR) condition at static magnetic field $\mu_0 H=27.2\ \si{mT}$ applied in an in-plane direction [see Fig. 1(b)].

Firstly, we generate the RF input field, $h_{1f}$, perpendicular to the static field, through the input CPW and the gold electrode which drives the initial magnetization precession in the sample at frequency 1\textit{f}. Next, the pump field, $h_{2f}$, that is parallel to static field, is generated from the short-ended CPW  [see Fig. 1(c)]. In the magnetic thin disk, the trajectory of the magnetization precession is distorted into an ellipse owing to the dynamical demagnetization field, which leads to the temporal change in the longitudinal component of magnetization with doubled frequency of ferromagnetic resonance, $M_{z,2f}$ [Fig. 1(d)]. Consequently, the pump field couples with the longitudinal component, leading to the parametric amplification of the precession dynamics \cite{Makiuchi2021,Brach2016,Hoshi2022}. The wavelength of the amplified spin waves is determined by dispersion relation [Fig. 1(e)]. The resulting precession can be detected via a.c. spin pumping and inverse spin hall effect (ISHE) as the voltage read-out from the sample measured by signal analyzer (SA)[Fig. 1(c)]. In this experiment, we set the frequency of the pump field  to be 2\textit{f} = 4.3 GHz and the input power to be $P_{1f}= 10\ \si{\micro W}$. 

\begin{figure}
\includegraphics{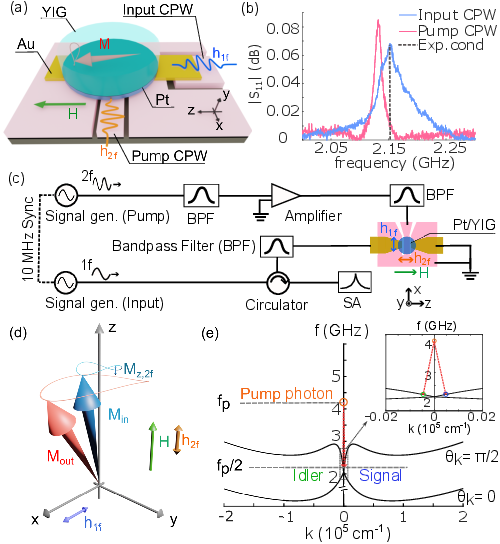}
\caption{\label{fig:epsart} (a) Sample configuration of YIG/Pt bilayer. (b) $|S_{11}|$ spectrum  taken at  $\mu_0H=27.2\ \si{mT}$ from different ports: Input CPW (blue) and Pump CPW (pink). (c) The microwave circuit diagram. (d) Schematic of elliptical trajectory of magnetisation precession in magnetic thin disk. $M_{\mathrm{in}}$ is the initial and $M_{\mathrm{out}}$ is the final magnetization. (e) Spin waves dispersion relation. A photon converts into two magnon with half of the photon frequency, $f_\mathrm{p}= \frac{\omega_{\mathrm{p}}}{2\pi}$.}
\end{figure}

To estimate the gain, we  distinguish the ISHE voltage representing the magnetization precession from the voltage that is simply reflected off the sample by subtracting the voltage from different field ($\mu_0 H=23\ \si{mT}$). We evaluate the gain as:
\begin{equation}
    \mathrm{Gain}= \frac{V_{2f}^{\mathrm{FMR}}-V_{2f}^{\mathrm{Non-FMR}}}{V_{1f}^{\mathrm{FMR}}-V_{1f}^{\mathrm{Non-FMR}}} ,			
\end{equation}
 where $V_{2f}^{\mathrm{FMR}}$ is the voltage read-out when both  $h_{2f}$ and $h_{1f}$ are applied, while $V_{1f}^{\mathrm{FMR}}$ is the read-out when only $h_{1f}$ is applied (pump field, $h_{2f}$ is switched off) at $\mu_0 H=27.2\ \si{mT}$. Lastly, $V_{2f}^{\mathrm{Non-FMR}}$ and $V_{1f}^{\mathrm{Non-FMR}}$ are the read-out voltages at $\mu_0 H=23\ \si{mT}$. 

We have performed the systematic measurement of the gain as a function of pump power and input phase, as shown in the heat map in Fig. 2(a). The measurement is taken below the pump threshold ($P_{2f}<P_{\mathrm{th}}$). The pump threshold is defined to be the critical pump power at which amplification takes place even in the absence of the input field to drive the initial magnetisation dynamics. 

In Fig. 2(b), we demonstrate the periodic dependence of the gain on the input phase, $\varphi_\mathrm{s}$. The period of the phase-dependence is $\pi$ with maximum gain (amplification) and minimum gain (attenuation) occurring every $\frac{\pi}{2}$ phase-shift of $h_{1f}$. 

\begin{figure*}[ht]
    \includegraphics{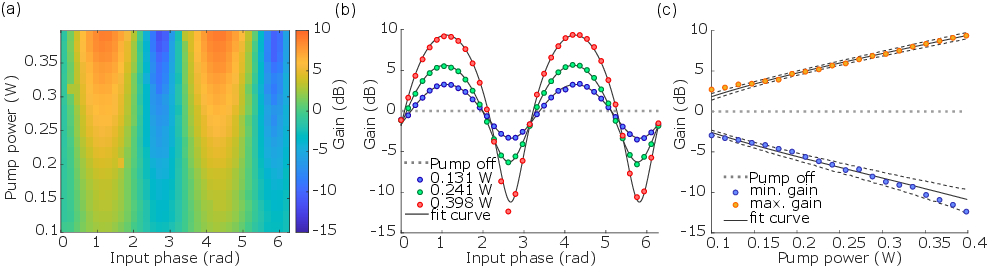}
    \caption{\label{fig:epsart} (a) Heat map of the gain (dB) with respect to pump power and input phase. (b) Gain dependence on input phases at three different pump power. (c) The dependence of maximum gain (amplification) and minimum gain (attenuation) on pump power. The black dashed-curves represent the 0.68 confidence bound of the fit curves. The grey dotted-line shows the gain when the pump field is switched off.}
\end{figure*}

To explain this phase-dependence of the gain, we consider that in a parametric amplification, a weak input signal wave and a strong pump wave are introduced into this non-linear system. The signal wave is then amplified and an idler wave, whose frequency is equal to the signal's frequency, is generated. Therefore, the resulting output wave is the superposition of the signal and idler waves. The two waves interfere with each other constructively or destructively, leading to amplification or attenuation of the signal which depends on the input signal phase.

To maximize the energy transferred from the pump into the signal and idler waves \cite{Brach2017,Lvov,Zakharov1975} and to meet the phase-matching condition (see Supplemental Material \cite{Suppl}), the phases of the signal ($\varphi_\mathrm{s}$), idler $(\varphi_\mathrm{i})$, and pump $(\varphi_\mathrm{p})$ waves must satisfy the relation  :
\begin{equation}
    \varphi_\mathrm{s}+\ \varphi_\mathrm{i}=\ \varphi_\mathrm{p}+\ \ \pi/2.
\end{equation}
From this relation, the phase difference between signal and idler, $\Delta \varphi_\mathrm{si}=\varphi_\mathrm{s} - \varphi_\mathrm{i}$,  in accordance to equation (1), follows:
\begin{equation}
    \Delta \varphi_\mathrm{si} = 2\Delta \varphi_\mathrm{sp} - \frac{\pi}{2},
\end{equation}
where the phase difference between the signal and pump,  $\Delta \varphi_\mathrm{sp}=\varphi_\mathrm{s} - \frac{\varphi_\mathrm{p}}{2}$.

Following equation (3), when $\Delta \varphi_{\mathrm{sp}} = \frac{\pi}{4}$ ($\frac{3\pi}{4}$), the signal-idler phase difference becomes  $\Delta \varphi_{\mathrm{si}} = 0$ ($\pi$) and they interfere constructively (destructively) (see Supplemental Material \cite{Suppl}). As a result, for a fixed pump's phase, we can expect an attenuation and amplification to occur every $\frac{\pi}{2}$ shift of the input signal's phase.

The dependence of the maximum gain and minimum gain on the amplitude of pump power is shown in Fig. 2(c). From the data, there are non-linear rise (decrease) of the maximum (minimum) gain as a function of pump power.

In order to discuss the dependence of the gain on the pump power and the input phase, we consider the microscopic picture of parallel parametric amplification process. Using the input-output formalism \cite{Walls, Yamamoto2016}, this process can be described by a system which interacts with the heat bath and the signal port, where the input and output signals are considered. The Hamiltonian describing this process is written as \cite{Yamamoto2016}:
\begin{equation}
    {\hat{H}}=  \hat{H}_{\mathrm{system}}+{\hat{H}}_{\mathrm{port}}+{\hat{H}}_{\mathrm{bath}}.
\end{equation}

The system Hamiltonian is expressed as \cite{Walls}
\begin{equation}
    \hat{H}_{\mathrm{system}}={\hat{H}}_0+{\hat{H}}_{\mathrm{int}}+{\hat{H}}_{\mathrm{pump}},
\end{equation}
where ${\hat{H}}_0$ is the Hamiltonian describing harmonic oscillators system, described as:
\begin{equation}
    {\hat{H}}_0=\ \hbar\omega{\hat{c}}^\dag{\hat{c}}+\hbar\omega_\mathrm{p}{\hat{a}}^\dag{\hat{a}},			
\end{equation}
where $\hat{c}$ ($\hat{c}^\dag$) is the annihilation (creation) operator of the magnon with frequency $\omega=2\pi f$ (after diagonalization of the quadratic terms in the dipolar interaction Hamiltonian \cite{Rez}). ${\hat{a}}$ ($\hat{a}^\dag$) is the annihilation (creation) operator of pump photon of frequency $\omega_\mathrm{p}$. In the degenerate parametric amplification case, we consider  $\omega_\mathrm{p}=2\omega$. 

The interaction Hamiltonian, ${\hat{H}}_{\mathrm{int}}$ describes the parametric process that splits a pump photon into two magnons; it is written as:
\begin{equation}
   {\hat{H}}_\mathrm{int}=\frac{i\hbar}{2}\rho_k\left({\hat{c}}^\dag{\hat{c}}^\dag{\hat{a}}-{\hat{c}}{\hat{c}}{\hat{a}}^\dag\right),	 
\end{equation}
where the coupling parameter between the pump photon and magnon, $\rho_k=\frac{\omega_M}{4} \sqrt{\frac{\omega_M}{\omega}}\sin^2{\theta_ke^{-i2\phi_k}}$; we use $\omega_M= \gamma \mu_0 M_\mathrm{s} $ with $M_s$ as the saturation magnetization, $\gamma$ as the gyromagnetic ratio, $\theta_k$ and $\phi_k$ as the polar and azimuthal angle of the wavevector, $\mathbf{k}$ \cite{Rez}. 

${\hat{H}}_{\mathrm{pump}}$ represents the driving pump fields in generating  the pump photons. It is written as 
\begin{equation}        {\hat{H}}_{\mathrm{pump}}=i\hbar\left(\varepsilon_\mathrm{p}e^{-i\left(\omega_\mathrm{p}t+\varphi_\mathrm{p}\right)}{\hat{a}}^\dag-{\varepsilon_\mathrm{p}e}^{i\left(\omega_\mathrm{p}t+\varphi_\mathrm{p}\right)}{\hat{a}}\right),
\end{equation}
where $\varphi_\mathrm{p}$ is the pump’s phase, $\varepsilon_\mathrm{p}$ is the amplitude of the classical pump field, $\varepsilon_\mathrm{p}= \gamma |h_{2f}|$ . 

The Hamiltonian describing the interaction between the magnon in the system and the input and output fields are expressed as
\begin{equation}     {\hat{H}}_{\mathrm{port}}=i\hbar\sqrt{\frac{\gamma_\mathrm{0}}{2\pi}}\left[\left({\hat{b}}_{\mathrm{in}}{\hat{c}}^\dag-{{\hat{b}}_{\mathrm{in}}^\dag{\hat{c}}}\right)+\left({\hat{b}}_{\mathrm{out}}{\hat{c}}^\dag-{{\hat{b}}_{\mathrm{out}}^\dag{\hat{c}}}\right)\right], 
\end{equation}
where ${\hat{b}}_{\mathrm{in}}$(${\hat{b}}_{\mathrm{in}}^\dag$) and ${\hat{b}}_{\mathrm{out}}$(${\hat{b}}_{\mathrm{out}}^\dag$) are the annihilation (creation) operators of the input and output field respectively. $\gamma_0$ is the damping constant between the system and the signal port. 

By input-output theory, the relation between the output and input operators to the magnon operator in the system are expressed as \cite{Yamamoto2016,Gardiner1985,Walls,jacobs2014}
\begin{equation}
    {\hat{b}}_{\mathrm{out}}={\hat{b}}_{\mathrm{in}} - \sqrt{2\pi\gamma_0}{\hat{c}}.
\end{equation}

Lastly the damping of photon and magnons to the phonons bath are assumed to be written as 
\begin{equation}   {\hat{H}}_{\mathrm{bath}}={\hat{c}}{\hat{\Gamma}}_1^\dag+{\hat{c}}^\dag{\hat{\Gamma}}_1+{\hat{a}}{\hat{\Gamma}}_2^\dag+{\hat{a}}^\dag{\hat{\Gamma}}_2,	
\end{equation}		
where ${\hat{\Gamma}}_i$$\left(\hat{\Gamma}_i^\dag \right)$ with $i=1,2$  are the annihilation (creation) operators of the harmonic oscillator in the thermal bath of phonons. 

We consider the coherent state of magnon mode $\left|c\right\rangle$ with eigenvalue $c$,  such that ${\hat{c}}\left|c\right\rangle=c\left|c\right\rangle$. Similarly, we have $\left|a\right\rangle$ with eigenvalue $a$
and ${\hat{a}}\left|a\right\rangle=a\left|a\right\rangle$ for the photon mode. Lastly, for the input and output operators, we have ${\hat{b}}_j\left|b_j\right\rangle=b_j\left|b_j\right\rangle$ ($j= \mathrm{in}, \mathrm{out}$), where $b_{\mathrm{in}}=\left|b_{\mathrm{in}}\right| e^{-i\varphi_\mathrm{s}}$ and $\varphi_\mathrm{s}$ is the phase of the input signal.  

In the rotating frame of frequency $\omega=\frac{\omega_\mathrm{p}}{2}$ , we derive the quantum Langevin equations by taking the trace over the reservoir to be
\begin{subequations}
    \begin{align}
        \begin{split}
            \frac{d c}{d t}=\sqrt{\frac{\gamma_\mathrm{0}}{2\pi}}\left(b_{\mathrm{in}}+b_{\mathrm{out}} \right) +\rho_kc^\ast a-\gamma_1c+\sqrt{\rho_k a}\eta_1,
        \end{split}\\
        \begin{split}
            \frac{d c^\ast}{dt}=\sqrt{\frac{\gamma_\mathrm{0}}{2\pi}}\left(b_{\mathrm{in}}^*+b_{\mathrm{out}}^* \right) +\rho_kca^\ast-\gamma_1c^\ast+\sqrt{\rho_k a^\ast}\eta_1^\dag,
        \end{split}\\
        \begin{split}
            \frac{d a}{d t}=\varepsilon_\mathrm{p}e^{-i\varphi_\mathrm{p}}-\frac{\rho_k}{2}c^2-\gamma_2a,
        \end{split}
    \end{align}
\end{subequations}
where $\gamma_1$ and $\gamma_2$ are the damping rate constants of the magnon and photon modes decay to phonon bath. $\eta_1\left(t\right)$ is the delta-correlated random force with zero mean (see Supplemental Material \cite{Suppl}).

By using the steady state solutions of equation (12) and considering substitutions with equation (10), we can get the expression of output field. The gain is then  defined as the square of ratio of output field when pump microwave ($h_{2f}$) is applied  $\left|b_\mathrm{out} (\varepsilon_{\mathrm{p}})\right|$ to its amplitude when the pump field is switched off $\left|b_\mathrm{out} (\varepsilon_{\mathrm{p}}= 0)\right|$. The gain is expressed as
\begin{equation}
        \begin{split}
              \mathrm{Gain} &= \left|\frac{b_\mathrm{out}(\varepsilon_{\mathrm{p}})}{b_\mathrm{out} (\varepsilon_{\mathrm{p}}= 0)}\right|^2 \\
     &=\beta \Big[\mathrm{X}_\mathrm{g}\cos^2{(\Delta \varphi_\mathrm{sp})} + \mathrm{Y}_\mathrm{g}\sin^2{(\Delta \varphi_\mathrm{sp})}\Big], 
     \end{split}
\end{equation}
where $\mathrm{X}_\mathrm{g}=\left(\frac{\gamma_0-\gamma_1+\tilde{\rho}\varepsilon_{\mathrm{p}}}{\gamma_0+\gamma_1-\tilde{\rho}\varepsilon_{\mathrm{p}}}\right)^2$,$\mathrm{Y}_\mathrm{g}=\left(\frac{\gamma_0-\gamma_1-\tilde{\rho}\varepsilon_{\mathrm{p}}}{\gamma_0+\gamma_1+\tilde{\rho}\varepsilon_\mathrm{p}}\right)^2$, and  $\beta = \left(\frac{\gamma_0+\gamma_1}{\gamma_0-\gamma_1}\right)^2$ (see Supplemental Material).
The phase difference between pump and signal is $\Delta\varphi_\mathrm{sp}=\varphi_\mathrm{s}-\frac{\varphi_\mathrm{p}}{2}$;  we use approximation, $|c| << |a| $, to substitute $a\approx\frac{\varepsilon_\mathrm{p}}{\gamma_2}e^{-i\varphi_\mathrm{p}}$; and $\tilde{\rho}=\frac{\rho_k}{\gamma_2}$ is the effective coupling constant between the magnon and the pump photon. The threshold  amplitude of pump field is $\varepsilon_{\mathrm{th}}=\frac{\gamma_0 + \gamma_1}{\tilde{\rho}}$.

Below threshold , $\varepsilon_\mathrm{p}<\varepsilon_{\mathrm{th}}$, there is a clear dependence of the gain on input phase $\varphi_s$ and pump field amplitude, $\varepsilon_\mathrm{p}$. Maximum gain (amplification) occurs when $\Delta\varphi_\mathrm{sp} = 0$, where $\mathrm{Gain}=\beta \mathrm{X}_\mathrm{g}=\beta \left(\frac{\gamma_0-\gamma_1+\tilde{\rho}\varepsilon_{\mathrm{p}}}{\gamma_0+\gamma_1-\tilde{\rho}\varepsilon_{\mathrm{p}}}\right)^2>1$ ; and minimum gain (attenuation) occurs when $\Delta\varphi_\mathrm{sp}=\pi/2$, where $ \mathrm{Gain}=\beta \mathrm{Y}_\mathrm{g}=\beta \left(\frac{\gamma_0-\gamma_1-\tilde{\rho}\varepsilon_{\mathrm{p}}}{\gamma_0+\gamma_1+\tilde{\rho}\varepsilon_\mathrm{p}}\right)^2<1$. 
For a fixed pump phase, $\varphi_\mathrm{p} $, these two conditions are alternatingly satisfied every time there is a phase-shift of $\frac{\pi}{2}$ of the input field.

\begin{figure}
    \includegraphics{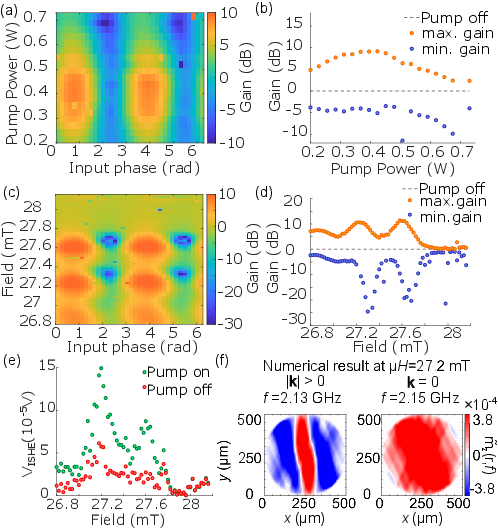}
    \caption{\label{fig:epsart} (a) Heat map of the gain with respect to pump power and input phase at high power range from 0.2 to 0.7 W. (b) The dependence of maximum gain (amplification) and minimum gain (attenuation) on pump power. (c) Heat map of the gain with respect to static magnetic field and input phase. Measurement is taken at pump power $P_{2f}=0.398\ \si{W}$. (d) The dependence of maximum gain (amplification) and minimum gain (attenuation) on static magnetic field. (e) The ISHE voltages at varying fields, taken at $1f= 2.15\ \si{GHz}$, with pump field (green) and without pump field (red). (f) Numerically calculated spatial patterns of standing spin waves at $\mu_0H=27.2\ $mT for $f= 2.13\ \si{GHz}$ and $f= 2.15\ \si{GHz}$.}
\end{figure}

Using equation (13), we can fit the data presented in Figs. 2(b) and (c). We note that there is a fixed offset of the phase ~10.5 rad, owing to the different pump and input channel in the experimental circuit. In Fig. 2(c), we consider pump power, $P_{2f}\propto{\varepsilon_\mathrm{p}}^2$ and threshold power, $P_{\mathrm{th}}\propto{\varepsilon_{\mathrm{th}}}^2$. From the fitting parameter, we can derive the threshold power at resonance condition  $\mu_0H=27.2\ $mT to be $P_{\mathrm{th}}=\left(0.8\pm0.1\right)\ \si{W} $, the damping constant of magnon mode to the heat bath, $\gamma_1=(1.5\pm0.1)\ \si{MHz}$, and the damping rate between the system and signal port to be $\gamma_0=(55.6\pm0.4)\ \si{MHz}$.  

On one hand, the damping rate  $\gamma_1$ is in the same order of magnitude as the ferromagnetic resonance (FMR) linewidth $\alpha f= 2.5\ \si{MHz}$, where we have evaluated the effective Gilbert damping parameter, $\alpha= 1.15 \times 10^{-3}$, from the field dependence of the linewidth of FMR (see Supplemental Material \cite{Suppl}); thus, verifying that $\gamma_1$ describes the rate of energy dissipation of the system to the lattice. On another hand, the damping rate $\gamma_0$ is consistent with the value of the linewidth of $|S_{11}|$ spectrum taken from the input CPW port with $\Delta f_\mathrm{Input} = 53.4\ \si{MHz}$ [see Fig. 1(b)]; and thus, affirming that $\gamma_0$ describes the energy loss between the system- YIG/Pt bilayer disk, to the signal port and the input CPW. 

Next, a systematic measurement has been performed at higher pump power range as shown in Fig. 3(a) and (b). The pump power is varied from 0.2 W to 0.7 W. In Fig. 3(a), we present the heat map of gain as a function of input phase and pump power. The alternating attenuation and amplification across the horizontal axis correspond to the periodic dependence of gain on the input phase. 

Figure 3(b) shows the maximum and minimum gain as a function of pump power. As we increase the pump power, the maximum gain rises and starts to saturate at around 0.4 W then it decreases at higher power. We found this saturation power to be lower than the  predicted threshold power $P_{\mathrm{th}}= (0.8 \pm 0.1) \si{W} $. The saturation of gain can be attributed to the Kerr non-linearity which limit the parametrically amplified magnon number\cite{Elyasi2022,YUAN2022,Zakharov1975,Lvov}. The magnon Kerr effect induces an appreciable shift of the magnon frequency\cite{Wang2016}, leading to a frequency mismatch between the magnon and the driving fields. Hence, increasing the pump power cannot amplify the magnon number any further. The subsequent decrease of the gain at higher pump power may be due to other non-linear effects that come from scatterings between different magnon modes, such as the second order Suhl instability which opens up an additional decay channel for the amplified magnons\cite{Rez1990,SUHL1957,Lvov}.

To search for the best condition in obtaining the largest gain, we measure the magnetic field dependence of the gain as shown in Figs. 3(c) and (d). The measurement is taken at pump power $P_{2f}=0.398\ \si{W}$. We observe two peaks structure at $\mu_0H=27.2\ \si{mT}$ with maximum gain of $ 10.7\ \si{dB}$ and at $\mu_0H=27.6\ \si{mT}$ with the gain of $11.4\ \si{dB}$ .  

We first present the ISHE voltage data in Fig. 3(e), defined as $V_{i}^{\mathrm{FMR}}-V_{i}^{\mathrm{Non-FMR}}$ where $i= 1f, 2f$, to explain the origin of the two peaks in the field dependence. When $h_{2f}$ is switched off (red markers), we observe a single peak at  $\mu_0H=27.2\ \si{mT}$ , corresponding to a uniform precession excited by ferromagnetic resonance (FMR) [ see also Fig. 1(b)]. However, upon applying $h_{2f}$ (green markers), we observe an additional peak at $\mu_0H=27.6\ \si{mT}$, which indicate an amplification of a different magnon mode.

One possible mode at $\mu_0H=27.6\ \si{mT}$  is the in-plane standing wave mode with $|\mathbf{k}| > 0$ \cite{Guo2014,Hioki12022}. In Fig. 1(b), we show the $|S_{11}|$ spectrum taken from the pump CPW (pink curve) at $\mu_0H=27.2\ \si{mT}$. The peak occurs at the frequency, $f = 2.13\ \si{GHz}$, less than that of the experimental condition at $f = 2.15\ \si{GHz}$ (grey dashed line). This is expected for a standing spin wave mode with $\mathbf{k} \parallel \mathbf{M} $  as frequency decreases with increasing wavevector. In consequence, at a fixed frequency of $f = 2.15\ \si{GHz}$, the excitation of $|\mathbf{k}| > 0$ standing wave mode will appear at a higher field corresponding to the second peak structure.

Figure 3(f) shows the numerically calculated spatial amplitude profiles of the two modes under the applied static field of $\mu_0H=27.2\ \si{mT}$. These results present the different standing waves modes: $|\mathbf{k}| > 0$ at $f = 2.13\ \si{GHz}$ and FMR mode at $f = 2.15\ \si{GHz}$, which are consistent to our interpretation of  the experimental data. The standing wave pattern at $f = 2.13\ \si{GHz}$ is slightly inclined due to presence of an anisotropy field that is directed at an angle to the external static field (see Supplemental Material \cite{Suppl}).  

From the experimental data and numerical calculations, we conclude that the two modes, FMR and $|\mathbf{k}| > 0$ standing wave, are the optimal conditions of parametric amplification  potentially due to the coupling efficiency between magnons and phonons, which is determined by coupling parameter and mode overlap. The coupling parameter between photon and magnon, $\rho_k=\frac{\omega_M}{4} \sqrt{\frac{\omega_M}{\omega}}\sin^2{\theta_ke^{-i2\phi_k}}$, is maximized at $\theta_k=\frac{\pi}{2}$ and at minimum frequency, $\omega$. As shown in the dispersion curve in Fig. 1(b), the FMR mode satisfies this condition. The coupling efficiency between the magnons in the YIG/Pt disk and pump photons in CPW is also affected by the mode overlaps between the magnons and photons \cite{Huebl2013,Zhang2014,Zhang2015}. As the CPW’s width ($100\ \si{\micro m}$) is narrower than the disk’s diameter ($500\ \si{\micro m}$), the maximum modes overlap occurs locally on their area of contact \cite{Zhang2014}. As we see in Fig. 3(f), the two modes, FMR and $|\mathbf{k}| > 0$ standing waves would cover the area of contact between the pumping CPW and the magnetic disk; thus, satisfying the second factor for the optimal amplification. 

In summary, we systematically investigate parallel parametric amplification and attenuation in YIG/Pt bilayer disk by means of ac spin pumping and inverse spin-Hall effect. The experimental data of the gain dependence on input phase and pump power well below the pump threshold are in good agreement with the theoretical analysis. Furthermore, we have measured the gain dependence on the pump power at higher power range and its dependence on the static magnetic field.

\begin{acknowledgments}
We thank T. Makiuchi and H. Shimizu for fruitful discussions. This work was partially supported by JST CREST (JPMJCR20C1 and JPMJCR20T2), JSPS KAKENHI (JP19H05600, JP22K14584, and JP22H05114), Advanced Technology Institute Research Grants 2022, Institute for AI and Beyond of the University of Tokyo, and IBM–UTokyo lab.
\end{acknowledgments}

\end{document}